\documentclass[sigconf,anonymous=False]{acmart}

\usepackage{times}
\usepackage{latexsym}
\usepackage{float}
\usepackage{amsmath}
\usepackage{placeins}
\usepackage{tabularx}
\usepackage{amssymb}
\usepackage{graphicx}
\usepackage{capt-of}
\usepackage{booktabs}
\usepackage{mathtools}
\usepackage{multirow}
\usepackage{subcaption}
\usepackage[inline]{enumitem}
\usepackage{url}
\usepackage{array}

\newcommand{\eg}{\emph{e.g.}}

\begin{document}
\title[Cross Domain Regularization for Neural Ranking Models]{Cross Domain Regularization for Neural Ranking Models using Adversarial Learning}

\setcopyright{acmcopyright}

\copyrightyear{2018} 
\acmYear{2018} 
\setcopyright{acmcopyright}
\acmConference[SIGIR '18]{The 41st International ACM SIGIR Conference on Research \& Development in Information Retrieval}{July 8--12, 2018}{Ann Arbor, MI, USA}
\acmBooktitle{SIGIR '18: The 41st International ACM SIGIR Conference on Research \& Development in Information Retrieval, July 8--12, 2018, Ann Arbor, MI, USA}
\acmPrice{15.00}
\acmDOI{10.1145/3209978.3210141}
\acmISBN{978-1-4503-5657-2/18/07}

\author{Daniel Cohen}
\affiliation{%
  Center for Intelligent Information Retrieval\\University of Massachusetts Amherst}
\authornote{Part of this work was done while Daniel Cohen was a research intern at Microsoft Research Cambridge.}
\email{dcohen@cs.umass.edu}

\author{Bhaskar Mitra, Katja Hofmann}
\affiliation{%
  Microsoft AI \& Research}
\email{bmitra@microsoft.com}
\email{katja.hofmann@microsoft.com}
\author{W. Bruce Croft}
\affiliation{
  Center for Intelligent Information Retrieval\\University of Massachusetts Amherst}
\email{croft@cs.umass.edu}


\renewcommand{\shortauthors}{D. Cohen et al.}

\begin{abstract}
Unlike traditional \emph{learning to rank} models that depend on hand-crafted features, neural representation learning models learn higher level features for the ranking task by training on large datasets. Their ability to learn new features directly from the data, however, may come at a price. Without any special supervision, these models learn relationships that may hold only in the domain from which the training data is sampled, and generalize poorly to domains not observed during training. We study the effectiveness of adversarial learning as a cross domain regularizer in the context of the ranking task. We use an adversarial discriminator and train our neural ranking model on a small set of domains. The discriminator provides a negative feedback signal to discourage the model from learning domain specific representations. Our experiments show consistently better performance on held out domains in the presence of the adversarial discriminator---sometimes up to 30\% on precision$@1$. 
\end{abstract}

%


\maketitle

\section{Introduction}
Several neural ranking models have been proposed recently that estimate the relevance of a document to a query by considering the raw query-document text \citep{moschitti} or based on the patterns of exact query term matches in the document \citep{Guo-DRMM}, or a combination of both \citep{mitra2017learning}. These models typically learn to distinguish between the input feature distributions corresponding to a relevant and a less relevant query-document pair by observing a large number of relevant and non-relevant samples during training. Unlike traditional \emph{learning to rank} (LTR) models that depend on hand-crafted features \citep{liu2009learning}, these deep neural models learn higher level representations useful for the target task directly from the data. Their ability to learn features from the training data is a powerful attribute that enables them to potentially discover new relationships not captured by hand-crafted features. However, as \citet{mitra2017introduction} discuss, the ability to learn new features may come at the cost of poor generalization and performance on domains not observed during training. The model, for example, may observe that certain pairs of phrases---\eg, ``Theresa May'' and ``Prime Minister''---co-occur together more often than others in the training corpus. Or, the model may conclude that it is more important to learn a good representation for ``Theresa May'' than for ``John Major'' based on their relative frequency of occurrences in training queries. While these correlations and distributions are important if our goal is to achieve the best performance on a single domain, the model must learn to be more robust to them if we instead care about ``out of box'' performance on unseen domains, \eg, older TREC collections \citep{voorhees2005trec}. In contrast, traditional retrieval models (\eg BM25 \citep{robertson2009probabilistic}) and LTR models based on aggregated count based features---that make fewer distributional assumptions---typically exhibit more robust cross domain performances.

Our goal is to train deep neural ranking models that learn useful representations from the data without ``overfitting'' to the distributions of the training domains. Recently, adversarial learning has been shown to be an effective cross domain regularizer suitable for classification tasks \citep{ganin2016domain, tzeng2017adversarial}. We adapt a similar strategy to force neural ranking models to learn more domain invariant representations. We train our neural ranking model on a small set of domains and evaluate its performance on held out domains. During training, we combine our ranking model with an adversarial discriminator that tries to predict the domain of the training sample based on the representations learned by the ranking model. The gradients from the adversarial components are reversed when backpropagating through the layers of the ranking model. This provides a negative feedback signal to the ranking model to discourage it from learning representations that may be significant only for specific domains. Our experiments show consistent improvements in ranking performance on held out domains from the proposed adversarial training---sometimes up to 30\% improvement on precision$@1$.

\begin{figure*}[t]
\center
\begin{subfigure}{0.30\textwidth}
    \includegraphics[width=\textwidth]{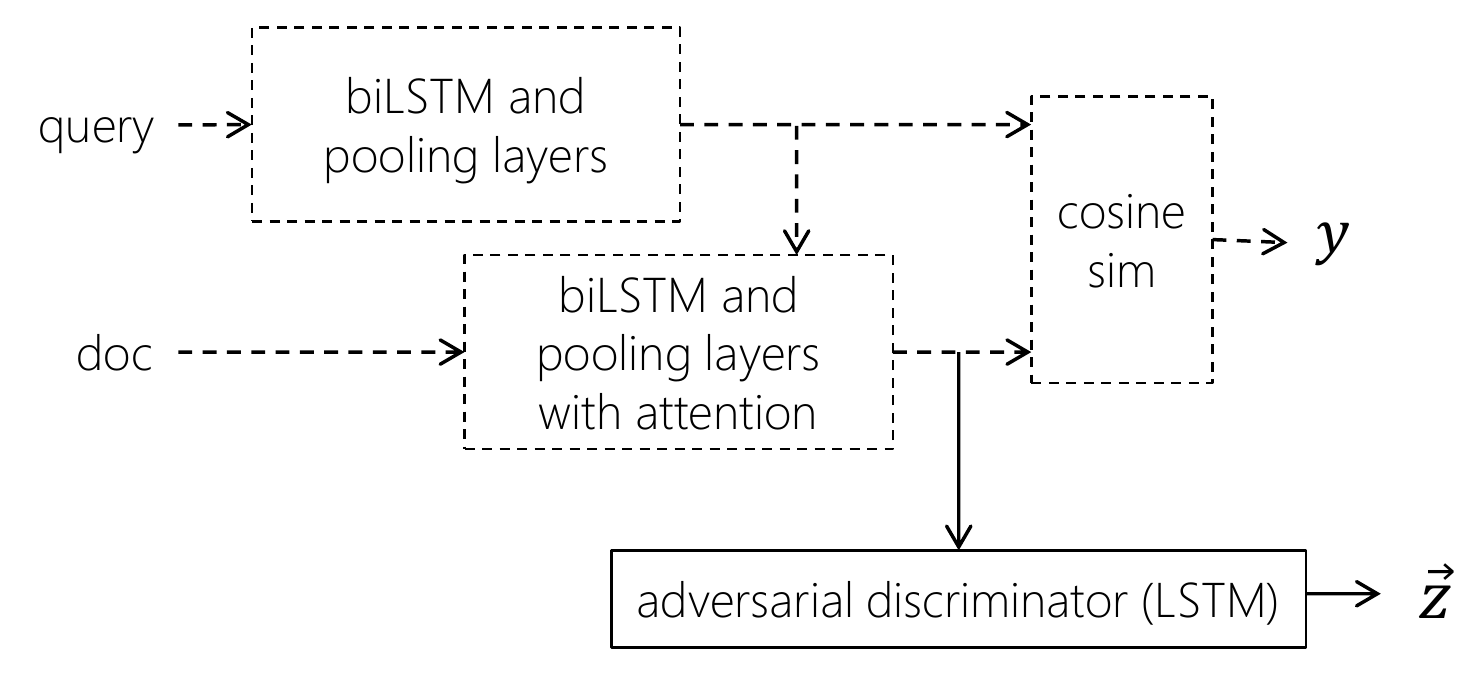}
    \caption{CosSim w/ adversarial discriminator}
    \label{fig:arch-cossim}
\end{subfigure}
\begin{subfigure}{0.55\textwidth}
    \includegraphics[width=\textwidth]{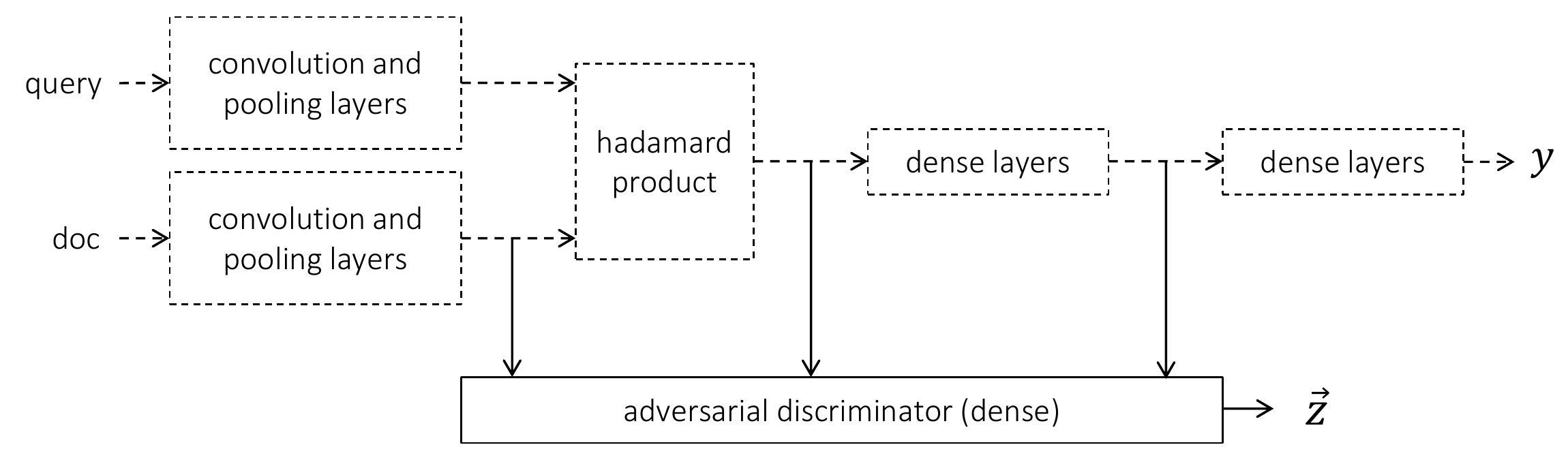}
    \caption{Duet-distributed w/ adversarial discriminator}
    \label{fig:arch-duet}
\end{subfigure}
\caption{Cross domain regularization of the two baseline models---CosSim and Duet-distributed---using an adversarial discriminator. The discriminator inspects the learned representations of the ranking model and provides a negative feedback signal for any representation that aids domain discrimination.}
\end{figure*}

\section{Related Work}
Adversarial networks surfaced shortly after they were introduced in the generative adversarial network (GAN) model. Goodfellow et al.~\cite{GAN-goodfellow} present a generative model that learns a distribution $p_G(x)$ that matches a true distribution $p_{data}(x)$.  The generative model receives training updates through a joint loss function shared with an adversarial network, the discriminator, that learns whether a sample is from $p_G(x)$ or $p_{data}(x)$ as a binary classification problem. The generator is penalized when the discriminator can successfully classify the sample origin, framing the relationship as a minimax game.
While initially proposed for generating continuous data, Donahue et al.~\cite{biGAN-donahue} extend this work by learning an encoder that maps the data to the latent space $\mathbf{z}$. They show that this can learn useful features for image classification tasks without the need for supervised training.\\
Tzeng et al.~\cite{tzeng2014deep} first propose a form of domain agnostic representation via \textit{domain confusion}, where the maximum mean discrepancy between the final layers of two identical networks over different domains is directly minimized. With the introduction of adversarial agents, Ganin et al.~\cite{ganin2016domain} approach the same task of domain agnostic representation by using an adversarial discriminator. The representation of the main network is forced away from a domain specific representation by reversing the gradient updates outside of the adversarial discriminator. \\
As previous methods used shared weights for both domains, Rozantsev et al.~\cite{rozantsev2016sharedDomainAdapt} expand on this work showing that unpairing a portion of the classification model, with only a small number of parameters shared prior to input into the final layers, can lead to effective adaptation in supervised and unsupervised settings. Recently, Tzeng et al.~\cite{tzeng2017adversarial} have represented a number of past domain adaptation works in a unified framework, referred to as \textit{Adversarial Discriminative Domain Adaptation}, that captures previous approaches as special cases and encompasses a GAN loss into the training of the classifier and adversarial discriminator. This methodology achieves robust domain agnostic models over computer vision collections.\\

\section{Cross domain regularization using adversarial learning}

The motivation of the adversarial discriminator is to force the neural model to learn domain independent features that are useful to estimate relevance. Conventional neural ranking models are trained to only optimize for relevance evaluations, disregarding the nature of features learned internally. We propose using an adversarial agent to force the features learned by the ranking model to be domain agnostic by shifting the model parameters in the opposite direction to domain specific spaces on the manifold. This cross domain regularization via domain confusion~\cite{tzeng2017adversarial} can be represented as a joint loss function:
\vspace{-0.2cm}
\begin{align}
\begin{split}
	\mathcal{L} &= \mathcal{L}_{\text{rel}}(q,doc_r,d_{nr},\theta_D,\theta_{\text{rel}}) \\ &+ \lambda\cdot\big(\mathcal{L}_{\text{adv}}(q,doc_r,\theta_D) + \mathcal{L}_{\text{adv}}(q,doc_{nr},\theta_D)\big)
\end{split}
\end{align}

where $\mathcal{L}_{\text{rel}}$ is a relevance based loss function and $L_\text{adv}$ is the adversarial discriminator loss. $q, doc_r$, and $doc_{nr}$ are the query, the relevant document, and the non-relevant documents, respectively. Finally, $\theta_{\text{rel}}$ and $\theta_D$ are the parameters for the relevance and the adversarial models, respectively. $\lambda$ determines how strongly the domain confusion loss should impact the optimization process. We treat it as a hyper-parameter in our training regime. The ranking model is trained on a set of train domains $D_\text{train} = \{d_1,\ldots,d_k\}$ separate from the set of held out domains $D_\text{test} = \{d_{k+1},\ldots,d_n\}$ on which it is evaluated.

The discriminator is a classifier that inspects the outputs of the hidden layers of the ranking model, and tries to predict the domain $d_\text{true} \in D_\text{train}$ of the training sample. The discriminator is trained using a standard cross-entropy loss.
\vspace{-0.2cm}
\begin{align}
\mathcal{L}_{\text{adv}}(q,doc,\theta_D) &= -\text{log}\big(p(d_\text{true}|q,doc,\theta_D)\big)\\
p(d_\text{true}|q,doc,\theta_D) &= \frac{exp(z_\text{true})}{\sum_{j \in D_\text{train}} exp(z_j)}
\end{align}

Gradient updates are performed via backpropagation through all subsequent layers, including those belonging to the ranking model. However, as proposed by Ganin et al.~\cite{ganin2016domain}, we utilize a gradient reversal layer. This layer transforms the standard gradient, $\frac{\delta L_\text{adv}}{\delta \theta}$ to its additive inverse, $-\frac{\delta \mathcal{L}_\text{adv}}{\delta \theta_{\text{rel}}}$. 
%
%
This results in $\theta_{\text{rel}}$ maximizing the domain identification loss, while still allowing $\theta_{\text{D}}$ to learn to discriminate domains. While not directly optimized, this can be viewed as modifying (1) via a sign change for $L_{\text{adv}}$.


\begin{table*}
\begin{tabular}{lc|cccc|cccccccc}
\toprule
    & & \multicolumn{4}{c}{CosSim} & \multicolumn{4}{c}{Duet-Dist.}\\
    \textit{source $\rightarrow$ target} & Size & \multicolumn{2}{c}{Original} & \multicolumn{2}{c}{Adv} & \multicolumn{2}{c}{Original} & \multicolumn{2}{c}{Adv}\\
    & & P@1 & MRR & P@1 & MRR & P@1 & MRR & P@1 & MRR\\
    \midrule
    All$\rightarrow$All& $142627$             &    \textbf{0.4229} & 0.6188 & 0.4213(-.3\%) & \textbf{0.6214}(+.4\%) & \textbf{0.4514} & \textbf{0.6136} & 0.4286(-5\%)$^\dagger$ & 0.6061(-1\%)$^\dagger$\\
    All*$\rightarrow$Sports & $139000$ & 0.3282 & 0.5194 & \textbf{0.4041}(+23\%)$^\dagger$ & \textbf{0.5925}(+12\%)$^\dagger$ & 0.2570 & 0.4567 &\textbf{0.3282}(+28\%)$^\dagger$ & \textbf{0.5011}(+10\%)$^\dagger$ \\
    Sports$\rightarrow$Sports & 3627        &  0.2146      & 0.5482 &  -     &- &0.2415 &0.3734 &- &- \\
    All*$\rightarrow$Home & $133372$   & 0.3460 & 0.5275 &\textbf{0.3645}(+5\%)$^\dagger$ & \textbf{0.5433}(+3\%)$^\dagger$ &0.3314 & 0.5285 &\textbf{0.3639}(+10\%)$^\dagger$ & \textbf{0.5457}(+3\%)$^\dagger$ \\
    Home$\rightarrow$Home & 9255            &  0.3014     & 0.5490 &    -   &- &0.2477 &0.4119 &- &- \\
    All*$\rightarrow$Politics & $138739$ & 0.3100 & 0.5101 &\textbf{0.3580}(+16\%)$^\dagger$ & \textbf{0.5507}(+8\%)$^\dagger$&0.3400 & 0.5291 &\textbf{0.3516}(+3\%)$^\dagger$ & \textbf{0.5342}(+3\%)$^\dagger$ \\
    Politics$\rightarrow$Politics & 3888    &  0.2219   & 0.5234 &   -    &- & 0.2160&0.5388 &- &- \\
    All*$\rightarrow$Travel & $140150$ & 0.2360 & 0.4486 &\textbf{0.2789}(+18\%)$^\dagger$ & \textbf{0.4723}(+5\%)$^\dagger$ &0.2158 & 0.4196 &\textbf{0.2842}(+32\%)$^\dagger$ & \textbf{0.4532}(+8\%)$^\dagger$ \\
    Travel$\rightarrow$Travel & 2477        &  0.2263       & 0.5181 &  -     &- &0.1895 & 0.3998& -&- \\
\bottomrule
\end{tabular}
\caption{Performance across L4 topics, where metrics under each collections represents the performance  of the model trained on the opposing two collections. \text{All*} is the entire L4 collection with target topic removed. $\dagger$ represents significance against non adversarial model ($p < 0.05$, Wilcoxon test)}
\label{tab:crossdomain}
\end{table*}

\paragraph{Passage Retrieval Models}
We evaluate our adversarial learning approach on the passage retrieval task. We employ the neural ranking model proposed by \citet{tan}---referred to as CosSim in the remaining sections---and the Duet model \cite{mitra2017learning} as our baselines. Our focus in this paper is on learning domain agnostic text representations. Therefore, similar to \citet{zamani2018neural} we only consider the distributed sub-network of the Duet model.

The CosSim model is an LSTM-based interaction focused architecture. We train the CosSim model in the same manner as~\cite{tan}, with a margin of 0.2 over a hinge loss function. The Duet-distributed is trained by maximizing the log likelihood of the correct passage, as originally proposed in~\cite{mitra2017learning}.
Similar to~\cite{nanni2017benchmark}, we adapt the hyper-parameters of the Duet model for passage retrieval. The output of the Hadamard product is significantly reduced by taking the max pooled representation, the query length is expanded to 20 from 8 tokens, and the max document length is reduced to 300 from the original 1000 tokens.

\bigskip\noindent
As opposed to past uses of adversarial approaches~\cite{ganin2016domain,CyCADA-Hoffman,tzeng2017adversarial}, ranking requires modeling an interaction between the query and the document. As shown in Figure~\ref{fig:arch-cossim}, the adversarial discriminator in our setting, therefore, inspects the joint query-document representation learned by the neural ranking models. For deeper architectures, such as the Duet-distributed, we allow the discriminator to inspect additional layers within the ranking model, as shown in Figure~\ref{fig:arch-duet}.

\section{Experiments}

\subsection{Data}
\label{sec:data}

\paragraph{\textit{L4}} We use Yahoo's Webscope L4 high quality "Manner" collection~\cite{yahoo}. For evaluation and training, all answers that were not the highest voted were removed from the collection to reduce label noise during training and provide a better judgment of performance during evaluation. Training, development, and test sets were created from a 80-10-10 split. Telescoping is used to create answer pools for evaluation from the top 10 BM25 retrieved answers as in~\cite{cohen-ictir}.

\indent \textit{InsuranceQA} In the InsuranceQA dataset, questions are created from real user submissions and the high quality answers come from insurance professionals. The dataset consists of 12,887 QA pairs for training, 1,000 pairs for validation, and two tests sets containing 1,800 pairs. For testing, each of the 1,800 QA pairs is evaluated with 499 randomly sampled candidate answers.

\indent\textit{WebAP} As both L4 and InsuranceQA are based on isolated passage retrieval for a directed question, we include the WebAP collection from Keikha et al.~\cite{webap} to examine how well a model trained on isolated passages with specific questions can generalize to a more general passage retrieval task. The format of this collection consists of 82 TREC queries with a total of 8,027 answer passages in total. As only relevant answer passages are annotated in this collection, we create non-relevant documents by using a sliding window of random size. Evaluation is done over a telescoped list of top 100 BM25 retrieved documents.

\begin{table*}
\begin{tabular}{l|cccc|cccccccc}
\toprule
    &  \multicolumn{4}{c}{CosSim} & \multicolumn{4}{c}{Duet-Dist.}\\
    \textit{source $\rightarrow$ target} & \multicolumn{2}{c}{Original} & \multicolumn{2}{c}{Adv} & \multicolumn{2}{c}{Original} & \multicolumn{2}{c}{Adv}\\
    &  P@1 & MRR & P@1 & MRR & P@1 & MRR & P@1 & MRR\\
    \midrule
    (InsuranceQA, L4)$\rightarrow$ WebAP    &    0.0901 & 0.2410 & \textbf{0.2500} & \textbf{0.3873} & 0.1250 & 0.4567 & \textbf{0.3286}$^\dagger$ & \textbf{0.5011}$^\dagger$\\
    
    (InsuranceQA, WebAP)$\rightarrow$ L4             &    0.1120 & 0.2957 & \textbf{0.2424}$^\dagger$ & \textbf{0.4335}$^\dagger$ & 0.0758 & 0.1939 & \textbf{0.3908}$^\dagger$ & \textbf{0.5642}$^\dagger$\\
    
    (L4, WebAP)$\rightarrow$ InsuranceQA              &   0.1406 & 0.4267 &   \textbf{0.1582} & \textbf{0.4717}$^\dagger$ & 0.0489  & 0.1473 & \textbf{0.1622}$\dagger$  & \textbf{0.3059}$\dagger$ \\
\bottomrule
\end{tabular}
\caption{Performance across collections, where metrics under each collections represents the performance  of the model trained on the opposing two collections. $\dagger$ represents significance against non adversarial model ($p < 0.05$, Wilcoxon test)}
\label{tab:crosscollection}
\vspace{-0.2cm}
\end{table*}

\subsection{Training}

We experimented with two different training settings---updating the ranking model and the discriminator parameters alternately as proposed by~\citet{GAN-goodfellow}, and simultaneously. We also tried different values for $\lambda$. Based on our validation results, we choose to train the CosSim model with alternate updates and $\lambda = 1$. For the Duet-distributed model, we see best performance with simultaneous updates and $\lambda = 0.25$. All models were trained with PyTorch~\footnote{https://github.com/pytorch/pytorch} and we implement early stopping based on the validation set.

\subsection{Evaluation}

We evaluate our proposed adversarial approach to cross domain regularization under two settings. Under the \emph{cross topic} setup, we consider the $25$ topics in the L4 dataset. We evaluate separately on four of these topics---Sports, Home, Politics, and Travel---each time training the corresponding models on the remaining 24 topics. For the \emph{cross collection} setup, we consider all three collections introduced in Section~\ref{sec:data}. Similar to the cross topic setting, we evaluate our models on each collection individually while training on the remaining two. However, due to more pronounced differences in both size and distributions between these collections---as compared to the differences between the L4 topics---our basic adversarial approach had limited success on the cross collection task. Thus, we adopt two additional changes to our training regime:
\begin{enumerate*}[label=(\roman*)]
    \item we sample the training data from the training collections equally to avoid over-fitting to any single collection, and
    \item we feed training samples from the evaluation collection to the adversarial discriminator.
\end{enumerate*}
We make sure that the training samples from the evaluation collection have no overlap with the test samples. In addition, we clarify that the ranking model receives no parameter updates from these training samples with respect to relevance judgments. These samples are only used to train the discriminator model's loss. This training setup may be appropriate when we want to train on some collections and evaluate on a different collection, where we can leverage the unlabeled documents from the target collection to at least guide the training of the adversarial component. 

\section{Results and Discussion}

\paragraph{Cross Topic} 

Table~\ref{tab:crossdomain} show the poor performance of the CosSim and Duet-distributed models on the four target topics when trained on the remaining collection. Notably, training on the topic specific data alone also performs poorly likely because of inadequate training data. However, in the presence of the adversarial discriminator both the models show significant improvement in performance on all held out topics. The improvements are somewhat bigger on the Duet-distributed baseline. We posit this is because the Duet-distributed model---with a deeper architecture---fits the training domain better at the cost of further loss in performance on the held out domains. Therefore, the adversarial learning has a stronger regularization opportunity on the Duet-distributed model.
\paragraph{Cross Collection} In similar vein as the cross topic evaluation, the incorporation of the adversarial signal significantly increases performance on the held out collections in Table~\ref{tab:crosscollection}. However, the difference in both size and distributional properties between these collections are far greater. Therefore, while the addition of the adversarial discriminator results in significant improvements---the absolute performance on the held out collections are still modest, even with adversarial regularization. We interpret these results as a reminder of the challenges in adapting these models to unseen domains.

\section{Conclusion and Future Work}

The proposed adversarial approach to cross domain regularization shows significant performance improvements consistently under two evaluation settings (cross topic and cross collection) and over  two different deep neural baselines. However, these improvements should be grounded in the realization that a model trained on large in-domain data is still likely to have a significant advantage over these models. Machine learning approaches to ad-hoc retrieval may need significantly more breakthroughs before achieving the level of robustness as some of the traditional retrieval models.

\section{Acknowledgements}
This work was supported in part by the Center for Intelligent Information Retrieval and in part by the Office of the Director of National Intelligence (ODNI), Intelligence Advanced Research Projects Activity (IARPA) via AFRL contract \#FA8650-17-C-9116 under subcontract \#94671240 from the University of Southern California. The views and conclusions contained herein are those of the authors and should not be interpreted as necessarily representing the official policies or endorsements, either expressed or implied, of the ODNI, IARPA, or the U.S. Government. The U.S. Government is authorized to reproduce and distribute reprints for Governmental purposes notwithstanding any copyright annotation thereon. Any opinions, findings and conclusions or recommendations expressed in this material are those of the authors and do not necessarily reflect those of the sponsor.

\FloatBarrier
\bibliographystyle{ACM-Reference-Format}
\bibliography{sample-bibliography} 


\begin{thebibliography}{20}


\ifx \showCODEN    \undefined \def \showCODEN     #1{\unskip}     \fi
\ifx \showDOI      \undefined \def \showDOI       #1{#1}\fi
\ifx \showISBNx    \undefined \def \showISBNx     #1{\unskip}     \fi
\ifx \showISBNxiii \undefined \def \showISBNxiii  #1{\unskip}     \fi
\ifx \showISSN     \undefined \def \showISSN      #1{\unskip}     \fi
\ifx \showLCCN     \undefined \def \showLCCN      #1{\unskip}     \fi
\ifx \shownote     \undefined \def \shownote      #1{#1}          \fi
\ifx \showarticletitle \undefined \def \showarticletitle #1{#1}   \fi
\ifx \showURL      \undefined \def \showURL       {\relax}        \fi
\providecommand\bibfield[2]{#2}
\providecommand\bibinfo[2]{#2}
\providecommand\natexlab[1]{#1}
\providecommand\showeprint[2][]{arXiv:#2}

\bibitem[\protect\citeauthoryear{Cohen and Croft}{Cohen and Croft}{[n. d.]}]%
        {cohen-ictir}
\bibfield{author}{\bibinfo{person}{Daniel Cohen} {and}
  \bibinfo{person}{W.~Bruce Croft}.} \bibinfo{year}{[n. d.]}\natexlab{}.
\newblock \showarticletitle{End to End Long Short Term Memory Networks for
  Non-Factoid Question Answering}. In \bibinfo{booktitle}{{\em ICTIR '16}}.
\newblock


\bibitem[\protect\citeauthoryear{Donahue, Kr{\"{a}}henb{\"{u}}hl, and
  Darrell}{Donahue et~al\mbox{.}}{2016}]%
        {biGAN-donahue}
\bibfield{author}{\bibinfo{person}{Jeff Donahue}, \bibinfo{person}{Philipp
  Kr{\"{a}}henb{\"{u}}hl}, {and} \bibinfo{person}{Trevor Darrell}.}
  \bibinfo{year}{2016}\natexlab{}.
\newblock \showarticletitle{Adversarial Feature Learning}.
\newblock \bibinfo{journal}{{\em CoRR\/}}  \bibinfo{volume}{abs/1605.09782}
  (\bibinfo{year}{2016}).
\newblock


\bibitem[\protect\citeauthoryear{Ganin, Ustinova, Ajakan, Germain, Larochelle,
  Laviolette, Marchand, and Lempitsky}{Ganin et~al\mbox{.}}{2016}]%
        {ganin2016domain}
\bibfield{author}{\bibinfo{person}{Yaroslav Ganin}, \bibinfo{person}{Evgeniya
  Ustinova}, \bibinfo{person}{Hana Ajakan}, \bibinfo{person}{Pascal Germain},
  \bibinfo{person}{Hugo Larochelle}, \bibinfo{person}{Fran{\c{c}}ois
  Laviolette}, \bibinfo{person}{Mario Marchand}, {and} \bibinfo{person}{Victor
  Lempitsky}.} \bibinfo{year}{2016}\natexlab{}.
\newblock \showarticletitle{Domain-adversarial training of neural networks}.
\newblock \bibinfo{journal}{{\em J. Mach. Learn. Res.\/}} \bibinfo{volume}{17},
  \bibinfo{number}{1} (\bibinfo{year}{2016}), \bibinfo{pages}{2096--2030}.
\newblock


\bibitem[\protect\citeauthoryear{Goodfellow, Pouget-Abadie, Mirza, Xu,
  Warde-Farley, Ozair, Courville, and Bengio}{Goodfellow et~al\mbox{.}}{2014}]%
        {GAN-goodfellow}
\bibfield{author}{\bibinfo{person}{Ian Goodfellow}, \bibinfo{person}{Jean
  Pouget-Abadie}, \bibinfo{person}{Mehdi Mirza}, \bibinfo{person}{Bing Xu},
  \bibinfo{person}{David Warde-Farley}, \bibinfo{person}{Sherjil Ozair},
  \bibinfo{person}{Aaron Courville}, {and} \bibinfo{person}{Yoshua Bengio}.}
  \bibinfo{year}{2014}\natexlab{}.
\newblock \showarticletitle{Generative Adversarial Nets}. In
  \bibinfo{booktitle}{{\em NIPS 2014}}. \bibinfo{publisher}{Curran Associates,
  Inc.}, \bibinfo{pages}{2672--2680}.
\newblock
\showURL{%
\url{http://papers.nips.cc/paper/5423-generative-adversarial-nets.pdf}}


\bibitem[\protect\citeauthoryear{Guo, Fan, Ai, and Croft}{Guo
  et~al\mbox{.}}{2016}]%
        {Guo-DRMM}
\bibfield{author}{\bibinfo{person}{Jiafeng Guo}, \bibinfo{person}{Yixing Fan},
  \bibinfo{person}{Qingyao Ai}, {and} \bibinfo{person}{W.~Bruce Croft}.}
  \bibinfo{year}{2016}\natexlab{}.
\newblock \showarticletitle{A Deep Relevance Matching Model for Ad-hoc
  Retrieval}. In \bibinfo{booktitle}{{\em CIKM '16}}. \bibinfo{publisher}{ACM},
  \bibinfo{address}{New York, NY, USA}, \bibinfo{pages}{55--64}.
\newblock
\showISBNx{978-1-4503-4073-1}
\showDOI{%
\url{https://doi.org/10.1145/2983323.2983769}}


\bibitem[\protect\citeauthoryear{Hoffman, Tzeng, Park, Zhu, Isola, Saenko,
  Efros, and Darrell}{Hoffman et~al\mbox{.}}{2017}]%
        {CyCADA-Hoffman}
\bibfield{author}{\bibinfo{person}{Judy Hoffman}, \bibinfo{person}{Eric Tzeng},
  \bibinfo{person}{Taesung Park}, \bibinfo{person}{Jun{-}Yan Zhu},
  \bibinfo{person}{Phillip Isola}, \bibinfo{person}{Kate Saenko},
  \bibinfo{person}{Alexei~A. Efros}, {and} \bibinfo{person}{Trevor Darrell}.}
  \bibinfo{year}{2017}\natexlab{}.
\newblock \showarticletitle{CyCADA: Cycle-Consistent Adversarial Domain
  Adaptation}.
\newblock \bibinfo{journal}{{\em CoRR\/}}  \bibinfo{volume}{abs/1711.03213}
  (\bibinfo{year}{2017}).
\newblock


\bibitem[\protect\citeauthoryear{Keikha, Park, Croft, and Sanderson}{Keikha
  et~al\mbox{.}}{2014}]%
        {webap}
\bibfield{author}{\bibinfo{person}{Mostafa Keikha}, \bibinfo{person}{Jae~Hyun
  Park}, \bibinfo{person}{W.~Bruce Croft}, {and} \bibinfo{person}{Mark
  Sanderson}.} \bibinfo{year}{2014}\natexlab{}.
\newblock \showarticletitle{Retrieving Passages and Finding Answers}. In
  \bibinfo{booktitle}{{\em ADCS '14}}. \bibinfo{publisher}{ACM},
  \bibinfo{address}{New York, NY, USA}, Article \bibinfo{articleno}{81},
  \bibinfo{numpages}{4}~pages.
\newblock
\showISBNx{978-1-4503-3000-8}
\showDOI{%
\url{https://doi.org/10.1145/2682862.2682877}}


\bibitem[\protect\citeauthoryear{Liu et~al\mbox{.}}{Liu et~al\mbox{.}}{2009}]%
        {liu2009learning}
\bibfield{author}{\bibinfo{person}{Tie-Yan Liu} {et~al\mbox{.}}}
  \bibinfo{year}{2009}\natexlab{}.
\newblock \showarticletitle{Learning to rank for information retrieval}.
\newblock \bibinfo{journal}{{\em Foundations and Trends{\textregistered} in
  IR\/}} \bibinfo{volume}{3}, \bibinfo{number}{3} (\bibinfo{year}{2009}),
  \bibinfo{pages}{225--331}.
\newblock


\bibitem[\protect\citeauthoryear{Mitra and Craswell}{Mitra and
  Craswell}{2018}]%
        {mitra2017introduction}
\bibfield{author}{\bibinfo{person}{Bhaskar Mitra} {and} \bibinfo{person}{Nick
  Craswell}.} \bibinfo{year}{2018}\natexlab{}.
\newblock \showarticletitle{An introduction to neural information retrieval}.
\newblock \bibinfo{journal}{{\em Foundations and Trends{\textregistered} in IR
  (to appear)\/}} (\bibinfo{year}{2018}).
\newblock


\bibitem[\protect\citeauthoryear{Mitra, Diaz, and Craswell}{Mitra
  et~al\mbox{.}}{2017}]%
        {mitra2017learning}
\bibfield{author}{\bibinfo{person}{Bhaskar Mitra}, \bibinfo{person}{Fernando
  Diaz}, {and} \bibinfo{person}{Nick Craswell}.}
  \bibinfo{year}{2017}\natexlab{}.
\newblock \showarticletitle{Learning to match using local and distributed
  representations of text for web search}. In \bibinfo{booktitle}{{\em WWW
  17}}. \bibinfo{pages}{1291--1299}.
\newblock


\bibitem[\protect\citeauthoryear{Nanni, Mitra, Magnusson, and Dietz}{Nanni
  et~al\mbox{.}}{2017}]%
        {nanni2017benchmark}
\bibfield{author}{\bibinfo{person}{Federico Nanni}, \bibinfo{person}{Bhaskar
  Mitra}, \bibinfo{person}{Matt Magnusson}, {and} \bibinfo{person}{Laura
  Dietz}.} \bibinfo{year}{2017}\natexlab{}.
\newblock \showarticletitle{Benchmark for complex answer retrieval}. In
  \bibinfo{booktitle}{{\em Proc. ICTIR}}. ACM, \bibinfo{pages}{293--296}.
\newblock


\bibitem[\protect\citeauthoryear{Robertson, Zaragoza, et~al\mbox{.}}{Robertson
  et~al\mbox{.}}{2009}]%
        {robertson2009probabilistic}
\bibfield{author}{\bibinfo{person}{Stephen Robertson}, \bibinfo{person}{Hugo
  Zaragoza}, {et~al\mbox{.}}} \bibinfo{year}{2009}\natexlab{}.
\newblock \showarticletitle{The probabilistic relevance framework: BM25 and
  beyond}.
\newblock \bibinfo{journal}{{\em Foundations and Trends{\textregistered} in
  IR\/}} \bibinfo{volume}{3}, \bibinfo{number}{4} (\bibinfo{year}{2009}),
  \bibinfo{pages}{333--389}.
\newblock


\bibitem[\protect\citeauthoryear{Rozantsev, Salzmann, and Fua}{Rozantsev
  et~al\mbox{.}}{2016}]%
        {rozantsev2016sharedDomainAdapt}
\bibfield{author}{\bibinfo{person}{Artem Rozantsev}, \bibinfo{person}{Mathieu
  Salzmann}, {and} \bibinfo{person}{Pascal Fua}.}
  \bibinfo{year}{2016}\natexlab{}.
\newblock \showarticletitle{Beyond Sharing Weights for Deep Domain Adaptation}.
\newblock \bibinfo{journal}{{\em CoRR\/}}  \bibinfo{volume}{abs/1603.06432}
  (\bibinfo{year}{2016}).
\newblock


\bibitem[\protect\citeauthoryear{Severyn and Moschitti}{Severyn and
  Moschitti}{2015}]%
        {moschitti}
\bibfield{author}{\bibinfo{person}{Aliaksei Severyn} {and}
  \bibinfo{person}{Alessandro Moschitti}.} \bibinfo{year}{2015}\natexlab{}.
\newblock \showarticletitle{Learning to Rank Short Text Pairs with
  Convolutional Deep Neural Networks}. In \bibinfo{booktitle}{{\em SIGIR}} {\em
  (\bibinfo{series}{SIGIR '15})}. \bibinfo{publisher}{ACM},
  \bibinfo{address}{New York, NY, USA}, \bibinfo{pages}{373--382}.
\newblock
\showISBNx{978-1-4503-3621-5}
\showDOI{%
\url{https://doi.org/10.1145/2766462.2767738}}


\bibitem[\protect\citeauthoryear{Surdeanu, Ciaramita, and Zaragoza}{Surdeanu
  et~al\mbox{.}}{2008}]%
        {yahoo}
\bibfield{author}{\bibinfo{person}{Mihai Surdeanu},
  \bibinfo{person}{Massimiliano Ciaramita}, {and} \bibinfo{person}{Hugo
  Zaragoza}.} \bibinfo{year}{2008}\natexlab{}.
\newblock \showarticletitle{Learning to rank answers on large online QA
  collections}. In \bibinfo{booktitle}{{\em ACL:HLT}}.
  \bibinfo{pages}{719--727}.
\newblock


\bibitem[\protect\citeauthoryear{Tan, Xiang, and Zhou}{Tan
  et~al\mbox{.}}{2015}]%
        {tan}
\bibfield{author}{\bibinfo{person}{Ming Tan}, \bibinfo{person}{Bing Xiang},
  {and} \bibinfo{person}{Bowen Zhou}.} \bibinfo{year}{2015}\natexlab{}.
\newblock \showarticletitle{LSTM-based Deep Learning Models for non-factoid
  answer selection}.
\newblock \bibinfo{journal}{{\em CoRR\/}}  \bibinfo{volume}{abs/1511.04108}
  (\bibinfo{year}{2015}).
\newblock
\showURL{%
\url{http://arxiv.org/abs/1511.04108}}


\bibitem[\protect\citeauthoryear{Tzeng, Hoffman, Saenko, and Darrell}{Tzeng
  et~al\mbox{.}}{2017}]%
        {tzeng2017adversarial}
\bibfield{author}{\bibinfo{person}{Eric Tzeng}, \bibinfo{person}{Judy Hoffman},
  \bibinfo{person}{Kate Saenko}, {and} \bibinfo{person}{Trevor Darrell}.}
  \bibinfo{year}{2017}\natexlab{}.
\newblock \showarticletitle{Adversarial discriminative domain adaptation}. In
  \bibinfo{booktitle}{{\em CVPR 17}}, Vol.~\bibinfo{volume}{1}.
  \bibinfo{pages}{4}.
\newblock


\bibitem[\protect\citeauthoryear{Tzeng, Hoffman, Zhang, Saenko, and
  Darrell}{Tzeng et~al\mbox{.}}{2014}]%
        {tzeng2014deep}
\bibfield{author}{\bibinfo{person}{Eric Tzeng}, \bibinfo{person}{Judy Hoffman},
  \bibinfo{person}{Ning Zhang}, \bibinfo{person}{Kate Saenko}, {and}
  \bibinfo{person}{Trevor Darrell}.} \bibinfo{year}{2014}\natexlab{}.
\newblock \showarticletitle{Deep domain confusion: Maximizing for domain
  invariance}.
\newblock \bibinfo{journal}{{\em arXiv preprint arXiv:1412.3474\/}}
  (\bibinfo{year}{2014}).
\newblock


\bibitem[\protect\citeauthoryear{Voorhees, Harman, et~al\mbox{.}}{Voorhees
  et~al\mbox{.}}{2005}]%
        {voorhees2005trec}
\bibfield{author}{\bibinfo{person}{Ellen~M Voorhees}, \bibinfo{person}{Donna~K
  Harman}, {et~al\mbox{.}}} \bibinfo{year}{2005}\natexlab{}.
\newblock \bibinfo{booktitle}{{\em TREC: Experiment and evaluation in
  information retrieval}}. Vol.~\bibinfo{volume}{1}.
\newblock \bibinfo{publisher}{MIT press Cambridge}.
\newblock


\bibitem[\protect\citeauthoryear{Zamani, Mitra, Song, Craswell, and
  Tiwary}{Zamani et~al\mbox{.}}{2018}]%
        {zamani2018neural}
\bibfield{author}{\bibinfo{person}{Hamed Zamani}, \bibinfo{person}{Bhaskar
  Mitra}, \bibinfo{person}{Xia Song}, \bibinfo{person}{Nick Craswell}, {and}
  \bibinfo{person}{Saurabh Tiwary}.} \bibinfo{year}{2018}\natexlab{}.
\newblock \showarticletitle{Neural ranking models with multiple document
  fields}. In \bibinfo{booktitle}{{\em Proc. WSDM}}. ACM,
  \bibinfo{pages}{700--708}.
\newblock


\end{thebibliography}

\end{document}